\documentclass[11pt]{article}
\begin{document}
\title{{\bf{\Large  Statistical interparticle potential on noncommutative space}}}
\author{
{\bf {\normalsize Sunandan Gangopadhyay}$^{a,b,c}
$\thanks{sunandan.gangopadhyay@gmail.com, sunandan@bose.res.in}},\,
{\bf {\normalsize Ram Narayan Deb}$^{d}
$\thanks{debram@rediffmail.com}}\\
{\bf {\normalsize Frederik. G. Scholtz}
$^{e,}$\thanks{fgs@sun.ac.za}}\\
$^{a}$ {\normalsize Department of Physics, West Bengal State University, Barasat, India}\\
$^{b}${\normalsize Visiting Associate in S.N. Bose National Centre for Basic Sciences,}\\
{\normalsize JD Block, Sector III, Salt Lake, Kolkata-700098, India}\\
$^{c}${\normalsize Visiting Associate in Inter University Centre for Astronomy and Astrophysics,}\\
{\normalsize Pune 411007, India}\\[0.1cm]
$^{d}${\normalsize Department of Physics, Krishnagar Government College,}\\
{\normalsize Krishnanagar, Nadia, West Bengal, Pin-741101, India}\\[0.1cm]
$^{e}$ {\normalsize National Institute of Theoretical Physics, 
Stellenbosch,}\\{\normalsize Stellenbosch 7600, South Africa}\\[0.3cm]
}
\date{}

\maketitle


\begin{abstract}

\noindent We compute the statistical potential between two particles in the coherent state
formalism on the deformed configuration space. 
The result obtained by using the coherent states having a further degree of freedom
(proposed in \cite{rohwer}) is compatible with Pauli's exclusion principle reassuring the
fact that noncommutative quantum mechanics may in some way encode the notion of physical extent
and structure.  


\\[0.3cm]

\end{abstract}



There seems to be growing consensus that our notion of spacetime has to be
drastically revised in a consistent formulation of quantum mechanics and gravity
\cite{dop, seib}. One possible generalization, suggested by string theory \cite{witten}, which 
has attracted much interest recently is that of noncommutative spacetime \cite{doug}.
This led to intense activities on the possible physical consequences of noncommutativity
in quantum mechanics \cite{mezin}-\cite{dulat}, quantum electrodynamics \cite{chaichian}, 
the standard model \cite{xcalmet}-\cite{reuter} and cosmology \cite{ramrez, branden}.

An interesting feature of noncommutative quantum mechanics and quantum field theories that has been observed
in the literature is that such theories admit a description in terms of extended objects. Indeed, it has been
shown \cite{susskind} that a free particle moving in the noncommutative plane can be thought of as two oppositely
charged particles interacting through a harmonic potential and moving in a strong magnetic field.
This demonstration therefore suggests that noncommutative quantum mechanics may in some way encode
the notion of physical extent and structure thereby admitting a description in terms of extended objects. 

The notion of spatial extent and structure within the formulation of quantum mechanics on noncommutative
plane has been explored recently \cite{rohwer} by introduction of coherent states over noncommutative
configuration space and the notion of a weak measurement as implemented through an appropriate positive
operator-valued measure. The essence of the approach is the introduction of position for the particle, which
can be done rigorously in the formalism of \cite{fgs}. It has been argued in \cite{rohwer} that this additional structure may
be interpreted as the physical extent. This was done by first considering the path integral representation
found in \cite{sg}.


In this paper, we shall study the interaction potential between two particles using coherent state
formalism developed in \cite{fgs}. We shall carry out our computations using the coherent states 
introduced in \cite{fgs} (satisfying a deformed completeness relation as we shall see below)
and then repeat the same exercise with the ones having a further degree of freedom
introduced in \cite{rohwer}.


To begin with, we present a brief review of the formulation developed in \cite{fgs}.
The first step is to define classical configuration space. In two dimensions, 
the coordinates of noncommutative configuration space satisfy the commutation relation 
\begin{eqnarray}
[\hat{x}, \hat{y}] = i\theta
\label{1}
\end{eqnarray} 
where without loss of generality it is assumed that $\theta$ is a real positive parameter. Using this, it is convenient to define the creation and annihilation operators
\begin{eqnarray}
\hat{b} = \frac{1}{\sqrt{2\theta}} (\hat{x}+i\hat{y})\quad,\quad
\hat{b}^\dagger =\frac{1}{\sqrt{2\theta}} (\hat{x}-i\hat{y})
\label{2}
\end{eqnarray}
that satisfy the Fock algebra $[ \hat{b}, \hat{b}^\dagger ] = 1$. 
The noncommutative configuration space is then 
isomorphic to the boson Fock space
\begin{eqnarray}
\mathcal{H}_c = \textrm{span}\{ |n\rangle\equiv 
\frac{1}{\sqrt{n!}}(\hat{b}^\dagger)^n |0\rangle\}_{n=0}^{n=\infty}
\label{3}
\end{eqnarray}
where the span is taken over the field of complex numbers.

\noindent The next step is to introduce the Hilbert space
of the noncommutative quantum system. We consider the set of Hilbert-Schmidt operators acting on noncommutative configuration space
\begin{equation}
\mathcal{H}_q = \left\{ \psi(\hat{x},\hat{y}): 
\psi(\hat{x},\hat{y})\in \mathcal{B}
\left(\mathcal{H}_c\right),\;
{\rm tr_c}(\psi^\dagger(\hat{x},\hat{y})
\psi(\hat{x},\hat{y})) < \infty \right\}.
\label{4}
\end{equation}
Here ${\rm tr_c}$ denotes the trace over noncommutative 
configuration space and $\mathcal{B}\left(\mathcal{H}_c\right)$ 
the set of bounded operators on $\mathcal{H}_c$. 
This space has a natural inner product and norm 
\begin{equation}
\left(\phi(\hat{x}, \hat{y}), \psi(\hat{x},\hat{y})\right) = 
{\rm tr_c}(\phi(\hat{x}, \hat{y})^\dagger\psi(\hat{x}, \hat{y}))
\label{inner}
\end{equation}
and forms a Hilbert space \cite{hol}. To distinguish it from the noncommutative configuration space $\mathcal{H}_c$, which is also a Hilbert space, we shall refer to it as quantum Hilbert space and use the subscripts $c$ and $q$ to make this distinction. Furthermore, we denote states in the noncommutative configuration space by $|\cdot\rangle$ and states in the quantum Hilbert space by $\psi(\hat{x},\hat{y})\equiv |\psi)$ and the elements of its dual (linear functionals) by $(\psi|$, which maps elements of $\mathcal{H}_q$ onto complex numbers by $\left(\phi|\psi\right)=\left(\phi,\psi\right)={\rm tr_c}\left(\phi(\hat{x}, \hat{y})^\dagger
\psi(\hat{x}, \hat{y})\right)$. 

\noindent Assuming commutative momenta, a unitary representation of the 
noncommutative Heisenberg algebra in terms of operators $\hat{X}$, $\hat{Y}$, $\hat{P}_x$ and $\hat{P}_y$ 
acting on the states of the quantum Hilbert space (\ref{4}) is easily found to be 
\begin{eqnarray}
\label{schnc}
\hat{X}\psi(\hat{x},\hat{y}) &=& \hat{x}\psi(\hat{x},\hat{y})\quad; \quad
\hat{Y}\psi(\hat{x},\hat{y}) = \hat{y}\psi(\hat{x},\hat{y})\nonumber\\
\hat{P}_x\psi(\hat{x},\hat{y}) &=& \frac{1}{\theta}[\hat{y},\psi(\hat{x},\hat{y})]= -i\frac{\partial\psi(\hat{x},\hat{y})}{\partial\hat{x}}\nonumber\\
\hat{P}_y\psi(\hat{x},\hat{y}) &=& -\frac{1}{\theta}[\hat{x},\psi(\hat{x},\hat{y})]= -i\frac{\partial\psi(\hat{x},\hat{y})}{\partial\hat{y}}~.
\label{action}
\end{eqnarray}
We use capital letters to distinguish operators acting on 
quantum Hilbert space from those acting on 
noncommutative configuration space. 

It should be mentioned that the more standard approach to noncommutative quantum field theory and noncommutative quantum mechanics is to deform the algebra of functions into a noncommutative algebra through the introduction of a, in general non-unique, star product.  The current approach turns out to be equivalent to this as was shown in \cite{gerhold} and \cite{fgs1}.  The present approach, however, offers a number of advantages.  In the context of noncommutative quantum field theories it provides a much more systematic and unambiguous derivation of the energy-momentum tensor \cite{gerhold}.  In the case of noncommutative quantum mechanics it gives greater clarity on the issue of choice of star product, e.g. Voros versus Moyal, as well as the physical interpretations associated with this choice \cite{fgs1}.  Furthermore, this is the only approach that has been used successfully to solve the noncommutative Schr\"{o}dinger equation for non-trivial potentials, i.e., apart from the free particle and harmonic oscillator.  Indeed, explicit solutions for the noncommutative well in two dimensions were constructed in \cite{fgs2} using this approach.   

The minimal uncertainty states in noncommutative 
configuration space, which is isomorphic to boson Fock space, 
are well known to be the normalized coherent states \cite{klaud}\footnote{Note that we restrict
ourselves only to the coherent states as they provide a particularly convenient basis.  However, they are not the only minimal uncertainty states.}
\begin{equation}
\label{cs} 
|z\rangle = e^{-z\bar{z}/2}e^{z b^{\dagger}} |0\rangle
\end{equation}
where, $z=\frac{1}{\sqrt{2\theta}}\left(x+iy\right)$ 
is a dimensionless complex number. These states provide an overcomplete 
basis in the noncommutative configuration space. 
Corresponding to these states we can construct a state 
(operator) in quantum Hilbert space as follows
\begin{equation}
|z, \bar{z} )=|z\rangle\langle z|.
\label{csqh}
\end{equation}
These states also have the property
\begin{equation}
\hat{B}|z, \bar{z})=z|z, \bar{z})\quad;\quad \hat{B}=\frac{1}{\sqrt{2\theta}}\left(\hat{X}+i\hat{Y}\right).
\label{p1}
\end{equation}
Writing the trace in terms of coherent states and using 
$|\langle z|w\rangle|^2=e^{-|z-w|^2}$ it is easy to see that 
\begin{equation}
(z, \bar{z}|w, \bar{w})=tr_{c}
(|z\rangle\langle z|w\rangle\langle w|)=
|\langle z|w\rangle|^2=e^{-|z-w|^2}
\label{p2}
\end{equation}
which shows that $|z, \bar{z})$ is indeed a Hilbert-Schmidt operator.  
We can now construct the `position' representation of a state 
$|\psi)=\psi(\hat{x},\hat{y})$ as
\begin{equation}
(z, \bar{z}|\psi)=tr_{c}
(|z\rangle\langle z| \psi(\hat{x},\hat{y}))=
\langle z|\psi(\hat{x},\hat{y})|z\rangle.
\label{posrep}
\end{equation}
In particular, introducing momentum eigenstates normalised such that $(p'|p)=\delta(p-p')$
\begin{eqnarray}
|p)&=&\sqrt{\frac{\theta}{2\pi\hbar^{2}}}e^{\frac{i}{\hbar}\sqrt{\frac{\theta}{2}}
(\bar{p}b+pb^\dagger)}\quad;\quad\hat{P}_{i}|p)=p_{i}|p)\\
p&=&p_{x}+ip_{y}\quad,\quad \bar{p}=p_{x}-ip_{y}\nonumber
\label{eg}
\end{eqnarray}
satisfying the completeness relation
\begin{eqnarray}
\int d^{2}p~|p)(p|=1_{q}
\label{eg5}
\end{eqnarray}
we observe that the wave-function of a free particle on the noncommutative plane is given by \cite{spal, fgs, sg, modak}
\begin{eqnarray}
(z, \bar{z}|p)=\sqrt{\frac{\theta}{2\pi\hbar^{2}}}
e^{-\frac{\theta}{4\hbar^{2}}\bar{p}p}e^{\frac{i}{\hbar}\sqrt{\frac{\theta}{2}}(p\bar{z}+\bar{p}z)}~.
\label{eg3}
\end{eqnarray}
The position eigenstates $|z,\bar{z})$, on the other hand satisfy the
following completeness relation
\begin{eqnarray}
\int \frac{dzd\bar{z}}{\pi}~|z, \bar{z})\star_{z}(z, \bar{z}|=1_{q}
\label{eg6}
\end{eqnarray}
where the star product between two functions 
$f(z, \bar{z})$ and $g(z, \bar{z})$ is defined as
\begin{eqnarray}
f(z, \bar{z})\star_{z} g(z, \bar{z})=f(z, \bar{z})
e^{\stackrel{\leftarrow}{\partial_{\bar{z}}}
\stackrel{\rightarrow}{\partial_z}} g(z, \bar{z})
\label{eg7}
\end{eqnarray}
with $\partial_z=\frac{\partial}{\partial z}$ and $\partial_{\bar{z}}=\frac{\partial}{\partial \bar{z}}$.
This implies that the operators
\begin{eqnarray}
\hat{\Pi}_{z}=\frac{1}{\pi\theta}|z, \bar z)e^{\stackrel{\leftarrow}{\partial_{\bar{z}}}
\stackrel{\rightarrow}{\partial_z}}(z, \bar z|\quad,\quad \int dx~dy~\hat{\Pi}_{z}=1_{q}
\label{oper}
\end{eqnarray}
provide an operator-valued measure in the sense \cite{bergou}. We can then give a consistent
probability interpretation by assigning the probability of finding the particle at position
$(x, y)$, given that the system is described by the pure state density matrix $\hat{\rho}=|\psi)(\psi|$,
to be 
\begin{eqnarray}
\label{probab}
P(z)&=&tr_{q}(\hat{\Pi}_{z}\hat{\rho})=(\psi|\hat{\Pi}_{z}|\psi)=\frac{1}{\pi\theta}(\psi|z, \bar z)
e^{\stackrel{\leftarrow}{\partial_{\bar{z}}}
\stackrel{\rightarrow}{\partial_z}}(z, \bar z|\psi)\\
\int dxdy~ P(z) &=&1 \nonumber
\end{eqnarray}
where $tr_{q}$ denotes the trace over the quantum Hilbert space.

\noindent With the above formalism in place, we now proceed to investigate the
statistical potential between two particles. Let us consider two particles in the
canonical ensemble described by a density matrix
\begin{eqnarray}
\hat\rho=\frac{1}{Z}e^{-\beta\hat{H}(\hat P)}\quad;\quad \beta=1/(k_B T)
\label{density}
\end{eqnarray}
where $\hat{H}(\hat P)$ is the free Hamiltonian and $Z=tr_{q}(e^{-\beta\hat{H}(\hat P)})$ is the partition function of the system.

\noindent The probability of finding the particles at positions $z_1$ and $z_2$ is now given by
\begin{eqnarray}
P(z_{1}, z_{2})&=&tr_{q}(\hat{\Pi}^{\pm}_{z_1, z_2}\hat{\rho})\nonumber\\
&=& \frac{1}{\pi^{2}\theta^2}tr_{q}(|z_1 , z_2 )_{\pm} \star_{z_1 ,z_2 ~\pm}(z_1 , z_2 |\hat\rho)                    
\label{prob1}
\end{eqnarray}
where the physical states $|z_1 , z_2 )_{\pm}$ is given by
\begin{eqnarray}
|z_1 , z_2 )_{\pm}=\frac{1}{\sqrt{2}}
\left\{|z_1, \bar{z}_1)|z_2 , \bar{z}_2)\pm |z_2 , \bar{z}_2)|z_1, \bar{z}_1)\right\}                   
\label{boson_states}
\end{eqnarray}
and the $+(-)$ signs stand for bosons(fermions).

\noindent To proceed further, we exploit the fact that the star product can be decomposed as
\begin{eqnarray}
e^{\stackrel{\leftarrow}{\partial_{\bar{z}}}
\stackrel{\rightarrow}{\partial_z}}=\int \frac{du d\bar{u}}{\pi}~e^{-|u|^2 + u\stackrel{\rightarrow}{\partial_z}                
+\bar{u}\stackrel{\leftarrow}{\partial_{\bar z}}}~.
\label{star}
\end{eqnarray}
We shall now use this form of the star product to compute $P(z_{1}, z_{2})$ in eq.(\ref{prob1}).
However, we must be careful to ensure that the derivatives
with respect to $z_1$ and $z_2$ only act on the ket $|z_1 , z_2 )_\pm$, whereas the derivatives with
respect to $\bar{z}_1$ and $\bar{z}_2$ only see the bra $_{\pm}(z_1 , z_2 |$. To ensure this we will temporarily let
$|z_1 , z_2 )_\pm \rightarrow |w_1 , w_2 )_\pm$ and then let 
$w_1 \rightarrow z_1$ and $w_2 \rightarrow z_2$ at the end of the computation.
Thus, we have
\begin{eqnarray}
P(z_{1}, z_{2})=\frac{1}{\pi^{2}\theta^2 Z}\lim_{w_1 \rightarrow z_1 , w_2 \rightarrow z_2}\int \frac{du d\bar{u}}{\pi}
\frac{dv d\bar{v}}{\pi}~                    
e^{-|u|^2 + u\stackrel{\rightarrow}{\partial_z}                
+\bar{u}\stackrel{\leftarrow}{\partial_{\bar z}}}
e^{-|v|^2 + v\stackrel{\rightarrow}{\partial_z}                
+\bar{v}\stackrel{\leftarrow}{\partial_{\bar z}}}\nonumber\\
\times~ _{\pm}(z_1 , z_2 |e^{-\beta\hat{H}}|w_1 , w_2 )_{\pm}\nonumber\\
=\frac{1}{\pi^{2}\theta^2 Z}\lim_{w_1 \rightarrow z_1 , w_2 \rightarrow z_2}\int \frac{du d\bar{u}}{\pi}\frac{dv d\bar{v}}{\pi}~              
e^{-|u|^2 + u\stackrel{\rightarrow}{\partial_z}                
+\bar{u}\stackrel{\leftarrow}{\partial_{\bar z}}}
e^{-|v|^2 + v\stackrel{\rightarrow}{\partial_z}                
+\bar{v}\stackrel{\leftarrow}{\partial_{\bar z}}}\nonumber\\
~~~~~~~~~~~~~~~~\times\left[(w_1, \bar{w}_1 |e^{-\beta\hat{H}}|z_1, \bar{z}_1)(w_2, \bar{w}_2|e^{-\beta\hat{H}}|z_2, \bar{z}_2)\pm (w_1, \bar{w}_1 |e^{-\beta\hat{H}}|z_2, \bar{z}_2)(w_2, \bar{w}_2 |e^{-\beta\hat{H}}|z_1, \bar{z}_1)\right]
\label{prob2}
\end{eqnarray}
where in the second line of the above equation, we have used the fact that for non-interacting particles, the Hamiltonian 
has the form of a sum of two single-particle Hamiltonians, i.e. $H = H_1 \otimes I_2 + I_1 \otimes H_2$. 

\noindent Introducing a complete set of momentum eigenstates (\ref{eg5}) and using eq.(\ref{eg3}), the above expression simplifies
\begin{eqnarray}
P(z_{1}, z_{2})&=&\frac{1}{\pi^2 Z}\left(\frac{1}{2\pi \hbar^2}\right)^{2}
\left[\left(\int d^{2}p~e^{-\beta H(p)}\right)^2  \right.\nonumber\\
&&\left. \pm\int d^{2}p d^{2}k~e^{-\beta(H(p) + H(k))}e^{-\frac{\theta}{2\hbar^2}|p-k|^2}
e^{\frac{i}{\hbar}\sqrt{\frac{\theta}{2}}\{(p-k)(\bar{z}_1 -\bar{z}_2)+(\bar{p}-\bar{k})(z_1 -z_2)\}}\right]
\label{prob3}
\end{eqnarray}
where $H(p)$ is the eigenvalue of the operator $\hat{H}(\hat P)$ in the eigenstate $|p)$.

\noindent Putting $H(p)=p^2/(2m)$ (the non-relativistic form of the Hamiltonian), we obtain
\begin{eqnarray}
P(z_{1}, z_{2})&=&\frac{1}{\pi^2 Z}\left(\frac{m}{\beta \hbar^2}\right)^{2}
\left[1 \pm \frac{\beta\hbar^2}{\beta\hbar^2 +2\theta m}\exp\left(-\frac{2\theta m}{\beta\hbar^2 +2\theta m}|z_1 - z_2|^2\right)\right]~.
\label{prob4}
\end{eqnarray}
The partition function $Z$ of the system can be obtained from the
normalization condition 
\begin{eqnarray}
\int dx_1 dy_1 dx_2 dy_2 ~ P(z_1, z_2) =1
\label{prob4a}
\end{eqnarray}
and reads in the limit of the area of the system $A\rightarrow\infty$
\begin{eqnarray}
Z=\frac{1}{\pi^2}\left(\frac{m}{\beta \hbar^2}\right)^2 A^2 ~.
\label{prob4b}
\end{eqnarray}
Substituting the value of $Z$ in eq.(\ref{prob4}), we have
\begin{eqnarray}
P(z_{1}, z_{2})&=&\frac{1}{A^2}
\left[1 \pm \frac{\beta\hbar^2}{\beta\hbar^2 +2\theta m}\exp\left(-\frac{2\theta m}{\beta\hbar^2 +2\theta m}|z_1 - z_2|^2\right)\right]~.
\label{prob4c}
\end{eqnarray}
Remarkably, the above result shows that there is a finite probability of two fermions sitting at the same place.
Similar results had been obtained earlier in the context of twist approach to 
noncommutative quantum field theory \cite{sgfgsbc, sgfgssaha}. However, in this paper we have not made use of this (twist)
approach to obtain the above result. The result is indeed a consequence of the 
coherent state formalism on the deformed configuration space (\ref{3}) developed in \cite{fgs}.



\noindent We now move on to carry out the same exercise using the coherent states (introduced in \cite{rohwer})
favouring the description of noncommutative quantum mechanics in terms of extended objects.
To proceed, let us consider the resolution of the identity (\ref{eg6}) on the quantum Hilbert space $\mathcal{H}_q$.
It is easy to see (by using the decomposition of the star product 
$\star_z = e^{\stackrel{\leftarrow}{\partial_{\bar{z}}}\stackrel{\rightarrow}{\partial_z}}$ (\ref{star})) that one can write down the identity operator $1_q$ in the quantum Hilbert space $\mathcal{H}_q$ as
\begin{eqnarray}
1_{q}&=&\int \frac{dzd\bar{z}}{\pi}~\int \frac{dv d\bar{v}}{\pi}~e^{-|v|^2}|z, \bar z) e^{v\stackrel{\rightarrow}{\partial_z}              
+\bar{v}\stackrel{\leftarrow}{\partial_{\bar z}}}(z, \bar z|\nonumber\\
&\equiv& \int\frac{dzd\bar{z}}{\pi}~\int \frac{dv d\bar{v}}{\pi}~|z, v)(z,v|
\label{extended}
\end{eqnarray}
where the states $|z, v)$ are defined as 
\begin{eqnarray}
|z, v)=e^{-\bar{v}v/2}e^{\bar{v}\partial_z}|z, \bar z) 
\label{new_states}
\end{eqnarray}
and are coherent states in $z$ with a further degree of freedom since they satisfy
\begin{eqnarray}
\hat{B}|z, v)=z|z, v) \quad, \quad \forall ~v.
\label{new_coherent}
\end{eqnarray}
From eq.(\ref{extended}) it is evident that the states $|z, v)$ form an overcomplete set in the
quantum Hilbert space $\mathcal{H}_q$. Consequently, one may associate a positive operator valued measure
with them, namely
\begin{eqnarray}
\hat{\Pi}_{z, v}=\frac{1}{\pi}|z, v)(z, v|
\label{povm}
\end{eqnarray}
as $\hat{\Pi}_{z, v}$ is positive and Hermitian on the quantum Hilbert space $\mathcal{H}_q$ and integrates
to the identity. As a result of this, one can once again define the corresponding probability distribution
in $z$ and $v$, assuming a pure-state density matrix $\hat{\rho}=|\psi)(\psi|$:
\begin{eqnarray}
P(z, v)=tr_{q}(\hat{\rho}\hat{\Pi}_{z, v})=(\psi|\hat{\Pi}_{z, v}|\psi).
\label{new_prob}
\end{eqnarray}
This probability gives us information not only about the position $z$ but also about a further
degree of freedom $v$, given that the system is prepared in state $\psi$. It is easy to see that
the probability of finding the particle localized at point $z$ (without detecting any information
regarding $z$) can be obtained from $P(z, v)$ by integrating over all $v$:
\begin{eqnarray}
P(z)=\int\frac{dv d\bar{v}}{\pi\theta}~P(z, v)
\label{old_prob}
\end{eqnarray}
with
\begin{eqnarray}
\hat\Pi_z=\int\frac{dv d\bar{v}}{\pi\theta}~\hat{\Pi}_{z, v}~.
\label{relation}
\end{eqnarray}
The physical interpretation of the additional degrees of freedom ($v$) has been discussed in \cite{rohwer}.

\noindent With the above formalism in hand, we now move on to evaluate the probability of finding two particles
at $(z_{1}, v_{1})$ and $(z_{2}, v_{2})$:
\begin{eqnarray}
P(z_{1}, v_{1}; z_{2}, v_{2})&=&tr_{q}(\hat{\Pi}^{\pm}_{z_1, v_1; z_2 , v_2}\hat{\rho})\nonumber\\
&=& \frac{N}{\pi^{2}}~_{\pm}(z_1 , v_1 ; z_2 ; v_2|e^{-\beta \hat{H}(\hat P)}|z_1 , v_1 ; z_2, v_2 )_{\pm}  \nonumber\\
&=& \frac{N}{\pi^{2}}\left[(z_1 , v_1 |e^{-\beta \hat{H}(\hat P)}|z_1 , v_1)(z_2 , v_2 |e^{-\beta \hat{H}(\hat P)}|z_2 , v_2)
\pm|(z_1 , v_1 |e^{-\beta \hat{H}(\hat P)}|z_2 , v_2)|^2\right] \nonumber\\                 
\label{new_prob_do}
\end{eqnarray}
where $N$ is some constant which has to be fixed from eq.(s)(\ref{old_prob}, \ref{prob4c}).

\noindent Introducing once again a complete set of momentum eigenstates (\ref{eg5}) and using the overlap \cite{rohwer}
\begin{eqnarray}
(z, v|p)=\sqrt{\frac{\theta}{2\pi\hbar^{2}}}
e^{-\frac{\theta}{4\hbar^{2}}\bar{p}p}e^{\frac{i}{\hbar}\sqrt{\frac{\theta}{2}}[p\bar{z}+\bar{p}(z+v)]}
e^{-\frac{1}{2}|v|^2}
\label{new_overlap}
\end{eqnarray}
we obtain
\begin{eqnarray}
P(z_{1}, v_{1}; z_{2}, v_{2})&=&N\left(\frac{\theta}{2G\pi \hbar^2}\right)^2
e^{-(1-\frac{\theta}{2G\hbar^2})(|v_1|^2 +|v_2|^2)}\nonumber\\
&&\times\left\{1\pm e^{-\frac{\theta}{G\hbar^2}|z_1 -z_2|^2}e^{-\frac{\theta}{2G\hbar^2}
\{|v_1 -v_2|^2 +(v_1 -v_2)(\bar{z}_1 -\bar{z}_2)+(\bar{v}_1 -\bar{v}_2)(z_1 -z_2)\}}\right\}
\label{prob_new}
\end{eqnarray}
where $G=\frac{\beta}{2m}+\frac{\theta}{2\hbar^2}$~. The above result clearly shows that there is a finite
probability of two fermions (having different degrees of freedom $v_1 \neq v_2$) sitting at the same place ($z_1 =z_2 $)
consistent with our earlier observation made from eq.(\ref{prob4c}). However, the probability vanishes when both
$z_1 =z_2 $ and $v_1 =v_2 $. This is indeed remarkable since the introduction of the new degrees of freedom is compatible with Pauli's exclusion principle. This is reassuring as this computation suggests that noncommutative quantum mechanics may in some
way admit a description in terms of extended objects \cite{rohwer}.

\section*{Acknowledgements} The authors would like to thank the referee for useful comments.


\end{document}